\documentclass[twocolumn,showpacs,preprintnumbers,amsmath,aps,amssymb]{revtex4}
\usepackage{graphicx}
\usepackage{dcolumn}
\usepackage{bm}

\begin{document}
\title{H-theorem and Thermodynamics for generalized entropies that depend only on the probability}
\author{Octavio Obreg\'on}
\email{octavio@fisica.ugto.mx}
\author{J. Torres-Arenas}
\email{jtorres@fisica.ugto.mx}
\author{A. Gil-Villegas }
\email{gil@fisica.ugto.mx}
\affiliation{Divisi\'on de Ciencias e Ingenier\'ias, Campus Le\'on, Universidad
de Guanajuato, Loma del Bosque 103, Fracc. Lomas del Campestre,  37150, Le\'on, Guanajuato, M\'exico.}%
\begin{abstract}

We consider a previously proposed non-extensive statistical mechanics in which the  entropy  depends only on the probability, this was obtained from a $f(\beta)$ distribution and its corresponding 
Boltzmann factor. We show that the first term correcting the usual entropy also arises from several $f(\beta)$  distributions,  we also construct the corresponding $H$ function and demonstrate that a 
 generalized $H$-theorem is fulfilled.  Furthermore, expressing this $H$ function as function of the simplest Maxwellian state we find, up to a first approximation some  modified thermodynamic quantities for an ideal gas. In order to gain some insight about the behavior of the proposed generalized statistics, we present some simulation results for the case of a square-well and Lennard-Jones potentials, showing that an effective repulsive interaction is obtained with the new formalism.

\end{abstract}
 \maketitle

\section{Introduction}

By considering non equilibrium systems with a long-term
stationary state that possess a spatio-temporally fluctuating
intensive quantity,  more general statistics can be formulated,
called Superstatistics \cite{1}. Selecting the temperature as  a
fluctuating quantity among various available intensive quantities;
in \cite{4} a formalism was developed  to deduce entropies
associated to the  Boltzmann factors $B(E)$ arising from their corresponding
assumed $f_q (\beta)$ distributions.   Following this
procedure, the Boltzmann-Gibbs entropy and the  Tsallis entropy  $S_q$
corresponding to the Gamma distribution $(\chi^2)$ and depending
on a constant parameter $q$, were obtained.  
 For the log-normal, $F$-distribution and other distributions it is not possible to get
closed analytic expression for their associated entropies and the
calculations were performed numerically utilizing the
corresponding $B(E)$ in each case.

All these  $f_q (\beta)$ distributions and their Boltzmann factors $B_q (E)$ obtained 
from  them, depend on a constant parameter $q$, actually the $F$-distribution 
also depends on a second constant parameter.  Consequently the associated entropies 
depend on $q$. An extensive discussion exists in the literature analyzing the possible 
viability of these kind of models to explain several physical phenomena 
 \cite{2,3}.

In previous works \cite{5,6} we proposed a generalized Gamma distribution depending on a parameter 
$p_l$ and calculated  its associated Boltzmann factor. We were able to find an entropy 
  that depends only on this parameter $p_l$.  By means of maximizing the  entropy, $p_l$ was
 identified with the probability distribution. Furthermore, by considering the corresponding generalization of the 
von Neumann entropy in \cite{7}  it was shown that this same modified von Neumann entropy can be found by means of a generalized 
Replica trick \cite{7,8}.

Since the fundamental results of Boltzmann \cite{Boltzmann} obtained in the frame of diluted gases, it is known that $H$-theorem is one of the cornerstones of Statistical Mechanics and Thermodynamics. Considering  a system 
of $N$ hard spheres all of the same size with no interaction among them he showed  that for the function  $H(t) = < \ln f >$, with $<>$ denoting the ensemble average,   it satisfies, $dH/dt \le 0$, which in the frame of a local or global 
equilibrium, encodes the first microscopic basis for the second law of thermodynamics.  In general, the function $H$ dictates the evolution of an arbitrary initial state for a gas into local equilibrium with 
the subsequent arrival to thermodynamic equilibrium. Several extensions to the H-theorem are known. A quantum version of the theorem was given by Pauli in the early 20's \cite{Pauli} and the first special relativistic version 
was presented by Marrot \cite{Marrot} with posterior modifications, within the special relativistic frame,  introduced by several authors \cite{Ehlers,Tauber,Chernikov}. More recently, efforts to develop an H-theorem that takes
 into account other characteristics like  frictional dissipation \cite{Bizarro}, leading to a modification of the classical non increase behavior for the $H$ function, or  a non-extensive quantum version 
\cite{Silva} imposing a restrictive interval for the $q$ parameter have been done.\\

In this work we follow the route of starting with a generalized  $H$ function which satisfies the $H$ theorem,  an entropy as a function of volume and temperature  
is obtained. Using this entropy, a broad thermodynamic information can be obtained. In the spirit of the original work of Boltzmann, an ideal gas is considered for the thermodynamical analysis. Some thermodynamic 
response functions are calculated presenting deviations with respect to conventional extensive quantities. Using the thermodynamic response functions and some approximations, we show 
that a modified non trivial  equation of state can be obtained. More yet, a universal correction function emerge from this analysis for all the thermodynamical quantities.


We will first , in Section II, propose  $f(\beta)$ distributions that do not depend on an arbitrary 
constant parameter, but instead on a parameter $p_l$ that can be identified with the probability associated with the 
microscopic configuration of the system  \cite{6}.  We will calculate the associated 
Boltzmann factors.  It will be shown that for small variance of the fluctuations a universal behavior is exhibited by these 
different statistics.  It should be noted that by changing in these $f(\beta)$ distribution $p_l$ by $-p_l$ another family of Boltzmann factors with the same 
correction terms arise but now alternating the signs in the correction terms.  We will not consider here these similar cases.  In section III, a relevant result is obtained;  in particular for the $H$ associated with the Gamma $(\chi^2)$ distribution we will 
show a   corresponding generalized $H$-theorem.

In section IV a calculation of the modified entropy, arising from the $H$-function, as a function of temperature and volume for an  ideal gas is given. Thermodynamic response functions  like heat capacity, and ratio of isothermal compressibility 
and thermal expansion coefficient are  calculated and relative deviations of the usual behavior are discussed. Finally, 
 some simulations results are given redifining the distribution probability using the generalized statistics. For the square-well and Lennard-Jones fluids, internal energies and heat capacities are given using both the standard Boltzmann-Gibbs statistics and generalized probability of this work.  Section V is devoted to present our conclusions.

\section{Generalized distributions and their associated Boltzmann factors}

We begin by assuming a Gamma (or $\chi^2$) distributed  inverse temperature
$\beta$ depending on $p_l$, a parameter to be identified with the  probability associated with the microscopic
configuration of the system by means of maximizing the associated entropy.  As the Boltzmann-factor is given by 

\begin{equation}
B(E) = \int f(\beta) e^{-\beta E} \ d\beta.  \label{1}
\end{equation} 

 We may write this parameter $p_l$
Gamma distribution as

\begin{equation}
f_{p_{l}}(\beta) = \frac{1}{\beta_0 p_l \Gamma\left(\frac{1}{p_l}\right)}
\left( \frac{\beta}{\beta_0}\frac{1}{p_l}\right)^{\frac{1-p_l}{p_l}}
e^{-\beta/\beta_0 p_l}, \label{2}
\end{equation}
where $\beta_0$ is the average inverse temperature.
Integration over $\beta$ yields the generalized Boltzman factor

\begin{equation}
B_{p_{l}}(E) = (1+ p_l \beta_0 E)^{- \frac{1}{p_l}}, \label{3}
\end{equation}
as shown in \cite{5}, this kind of expression can be expanded for small
$p_l\beta_0E$,  to get

\begin{equation}
B_{p_l}(E) =    e^{-\beta_0 E} \left[1+ \frac{1}{2}p_l \beta^2_0 E^2 - \frac{1}{3} p^2_l \beta^3_0 E^3 + ...\right].\label{4}
\end{equation}

We follow now the same procedure for the log-normal distribution, this  can be written in terms of $p_l$ as

\begin{equation}
f_{p_{l}}(\beta) = \frac{1}{\sqrt{2\pi} \beta [ \ln ( p_l+1)]^{1/2}}  \exp
\{ - \frac{ \left[ \ln  \frac{\beta (p_l+1)^{1/2}}{\beta_0} \right]^2}{2 \ln (p_l+1)} \}, \label{5}
\end{equation}
the generalized Boltzmann factor  can be obtained to leading order, for small variance of the
inverse temperature fluctuations,

\begin{equation}
B_{p_{l}}(E) = e^{-\beta_0 E} \left[1 + \frac{1}{2} p_l \beta^2_0 E^2 - \frac{1}{6} p^2_l(p_l+3)\beta^3_0 E^3+ \cdots \right]. \label{6}
\end{equation}

In general, the $F$-distribution has two free constant parameters. We consider,
particularly, the case in which one of these constant parameters is chosen as $v=4$. For this 
value of the constant parameter we define a $F$-distribution in function of the inverse of the
temperature and $p_l$ as

\begin{equation}
f_{p_{l}}(\beta)= \frac{\Gamma\left(\frac{8p_l-1}{2p_l-1}\right)}{\Gamma\left((\frac{4p_l+1}{2p_-1}\right)}
\frac{1}{\beta_0^2}
\left(\frac{2p_l-1}{p_l+1}\right)^2\frac{\beta} {\left(1+\frac{\beta}{\beta_0}\frac{2p_l-1}{p_l+1}\right)^{\left(
\frac{8p_l-1}{2p_l-1}\right)}},  \label{7}
\end{equation}
 once more the associated Boltzmann factor   can not be evaluated
in a closed form, but for small variance of the fluctuations we obtain the series expansion

\begin{equation}
B_{p_{l}}(E) = e^{-\beta_0E} \left[ 1+ \frac{1}{2}p_l \beta^2_0 E^2 + \frac{1}{3} p_l \frac{(5 p_l-1)}{p_l-2} \beta^3_0 E^3 + ... \right]. \label{8}
\end{equation}

As shown in  \cite{5,6} one can obtain in a closed form the entropy corresponding to  (Eqs. \ref{2}, \ref{3}) resulting in 

\begin{equation}
S = k \displaystyle\sum_{l=1}^{\Omega} (1-p_l^{p_l}),\label{9}
\end{equation}
where $k$ is the conventional constant and $\displaystyle\sum_{l=1}^{\Omega} p_l=1$.  The expansion of (Eq. \ref{9})  gives

\begin{equation}
-\frac{S}{k}= \displaystyle\sum_{l=1}^{\Omega} p_l \ln{p_l} +
\frac{(p_l \ln{p_l})^2}{2!} + \frac{(p_l\ln{p_l)}^3}{3!}+\cdots . \label{10}
\end{equation}

Given that the Boltzmann factors (Eqs. 4,6,8) coincide up to the second term for the Gamma $(\chi^2)$, log-normal and $F$-distributions, for enough small 
$p_l\beta_0E$ the entropy (Eq. \ref{10}) correspond to all these distributions up to the first term that modifies the usual entropy.  
We expect  at least this modification to the entropy for several possible $f(\beta)$ distributions.  

\section{A generalized H-theorem}
The usual H-theorem is established for the $H$ function defined as
\begin{equation}
H = \int d^3r d^3 p f \ln f.
\end{equation}
The essential of this theorem is to ensure that for any initial state, a gas that satisfies Boltzmann equation approach to a local equilibrium state, which means $dH/dt \le 0$.

The new $H$ function can be written as
\begin{equation}
H = \int d^3r d^3p ( f^f -1).
\end{equation}
Considering the partial time derivative (because in general the gas is not homogeneous), we have
\begin{equation}
\frac{\partial H}{\partial t} = \int d^3rd^3p [   \ln f + 1 ] e^{f \ \ln f} \ \frac{\partial f}{\partial t}.
\end{equation}
Using the mean value theorem for integrals, and realizing that the factor $e^{f \ \ln f}$ is always positive, it follows from the conventional H-theorem that the variation of the new H-function with time satisfies
\begin{equation}
\frac{\partial H}{\partial t}  \le 0 .
\end{equation}
This is a very interesting result, we can see that other possible generalizations, for which the multiplying factor appearing in the integral is positive defined, will preserve the corresponding H-theorem.\\

\section{Thermodynamics}  
It is well known that Boltzmann equation  under very general considerations admits solutions with global existence and exponential or polynomial decays to Maxwellian states \cite{Gressman}. The simplest Maxwellian state among the five-parameter family is given by

\begin{equation}
f(\vec{p}) =  n \Lambda^3 \ e^{\frac{- (\vec{p} - \vec{p}_0)^2}{2mkT}},
\end{equation} 
where  $n=N/V$, $\Lambda = \frac{h}{\sqrt{2\pi mkT}}$, with $h$ the Planck's constant. Since this state is reached by the system under very general conditions, we propose  the new H function defined as

\begin{equation}
H = \int \ d^3rd^3p \ \left( e^{f \ln f} - 1 \right),
\end{equation}
by expanding and doing the corresponding integration, we obtain the exact relation
\begin{equation}
H =  \sum_{l=1}^{\infty} \sum_{m=0}^l    \frac{(-1)^m  (2m+1)! }{(l-m)!  l^{m+ 3/2} (m!)^2 2^{2m}}   n^l \Lambda^{3(l-1)}   \left( \ln n \Lambda^3 \right)^{l-m} .
\end{equation}
In order to get an insight into the new contributions coming from the generalized H-function to the thermodynamics of a system, we keep only the first terms of the previous series

\begin{eqnarray}
H & =& n ( \ln n \Lambda^3 - \frac{3}{2}) + \frac{n^2 \Lambda^3}{2^{5/2} } \nonumber \\
& & \left[ \left(  \ln n \Lambda^3\right)^2 - \frac{3}{2} \ln n \Lambda^3 + \frac{15}{16} \right].
\label{H0} 
\end{eqnarray}
In the classical limit and for the system  not far from equilibrium we have the proportionality between the $H$ function and entropy
\begin{equation}
H = - \frac{S}{k },
\end{equation}
from which expression (\ref{H0} ) multiplied by $-k$ gives the entropy (to the corresponding order)
where the ideal contribution is given by $S_{ideal} =  -kN ( \ln n \Lambda^3 - \frac{3}{2}) $ and the rest of the  terms correspond to the corrections to the thermodynamics of the system. The extensivity property is broken by these new terms. We notice that the Sackur-Tetrode expression for the entropy of the ideal gas $S = kN\ln{v} + \frac{3kN}{2}\ln{\frac{u}{b}} + \frac{5kN}{2}$ 
 can be recovered by an ad-hoc fixing term as it was originally proposed by Gibbs. 

With the knowledge of the entropy as a function of volume and temperature it is possible to calculate some of the response functions for the system

\begin{equation}
\left( \frac{\partial S}{\partial T} \right)_V = \frac{1}{T} C_V, \qquad \left( \frac{\partial S}{\partial V} \right)_T = \frac{\alpha}{\kappa_T},
\end{equation}

\begin{equation}
\alpha = \frac{1}{V} \left( \frac{\partial V}{\partial T} \right)_P , \qquad  \kappa_T =  - \frac{1}{V}\left( \frac{\partial V}{\partial P} \right)_T ,
\end{equation}
where $\alpha$ is the thermal expansion coefficient and $\kappa_T$ is the isothermal compressibility.
For an ideal gas $\alpha = 1/T$, $\kappa_T = 1/P$, therefore, using the equation of state $\alpha/\kappa_T = k_B n$, where $k_B$ is the Boltzmann' constant and $n$ is the number of moles. We obtain

\begin{equation}
\frac{C_V}{C_V^{ideal}} \sim 1 + \frac{n \Lambda^3}{2^{5/2} } \left[  \left(  \ln n \Lambda^3\right)^2 + \frac{1}{2} \ln  n \Lambda^3  - \frac{9}{16}\right] ,
 \end{equation}

\begin{equation}
\frac{\left(\frac{\alpha}{\kappa_T}\right)}{\left(\frac{\alpha}{\kappa_T}\right)_{ideal}} \sim 1 +  \frac{n \Lambda^3}{2^{5/2} } \left[  \left(  \ln n \Lambda^3 \right)^2 + \frac{1}{2} \ln n \Lambda^3  - \frac{9}{16}\right] ,
\end{equation}
The first contribution gives us the well known  result for a classical ideal gas. It is remarkable that for both quantities the relative deviation from ideality have the same functional form. In fact, such deviation can be expressed in terms of only one variable (see Figure 1), $x \equiv n \Lambda^3$
\begin{equation}
F(x) \equiv \frac{x}{2^{5/2} } \left[  \left(  \ln x \right)^2 + \frac{1}{2} \ln x  - \frac{9}{16}\right] 
\end{equation}
It can be observed that the usual behavior is obtained for very low densities or very high temperatures;  deviations are expected for very low temperatures or very high densities. In particular, high densities could be achieved with very small volumes (for fixed $N$), i.e., for confined systems.\\

\begin{figure}[htb]
  \centering
    \includegraphics[scale=0.3]{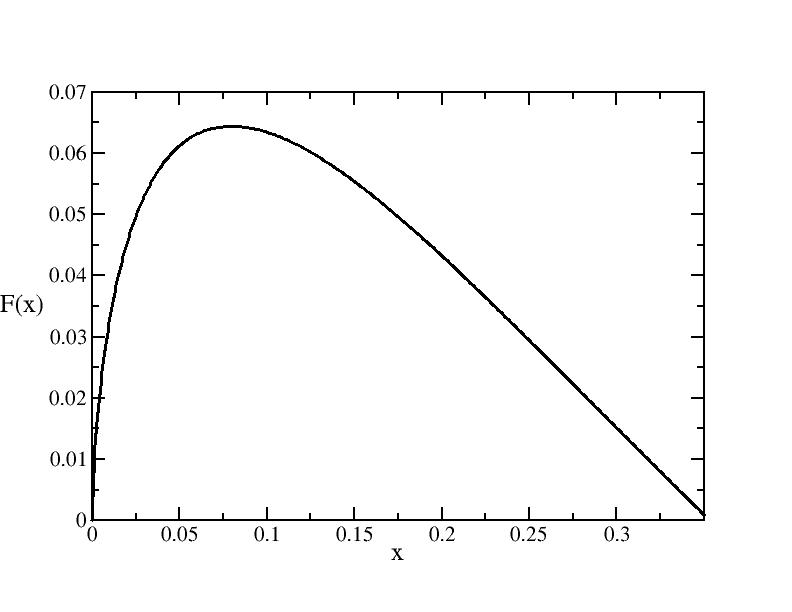}
\caption{Relative deviation from ideality for the thermodynamic quantities of the system. Ideal behavior is recovered for values of this function equal to zero.}
\label{fig:1}
\end{figure}

It is a well known thermodynamic result that an equation of state (pressure as a function of volume and temperature)  can be obtained if  the response functions $\alpha$ and $\kappa_T$ are given 
\begin{equation}
dP = \frac{\alpha}{\kappa_T} dT - \frac{1}{V\kappa_T} dV,
\end{equation}
in terms of the new variables $(n, \Lambda)$ the change in pressure can be written as
\begin{equation}
dP = -\frac{h^2\Lambda^{-3}}{\pi m k} \frac{\alpha}{\kappa_T} d\Lambda + \frac{1}{n \kappa_T} dn,
\end{equation}
and the response functions $\alpha , \kappa_T$ must be expressed in terms of $n,\Lambda$.\\
The quotient $\alpha/\kappa_T$ is known exactly, but the compressibility is not. It is possible to obtain an approximate equation of state  considering an
isochoric process, for it, an exact expression can be given (up to the considered order in our treatment).  Even within this coarse approximation,  a remarkable
similar expression to the previous obtained for the heat capacity and the expansion coeffcient is found. We can conjecture that the exact relative deviation for the pressure and the rest of the thermodynamic variables  is given by the function $F(x)$ up to the considered order in the expansion resembling the universal behavior found for the distribution functions .\\

\begin{figure}[htb]
  \centering
    \includegraphics[scale=0.3]{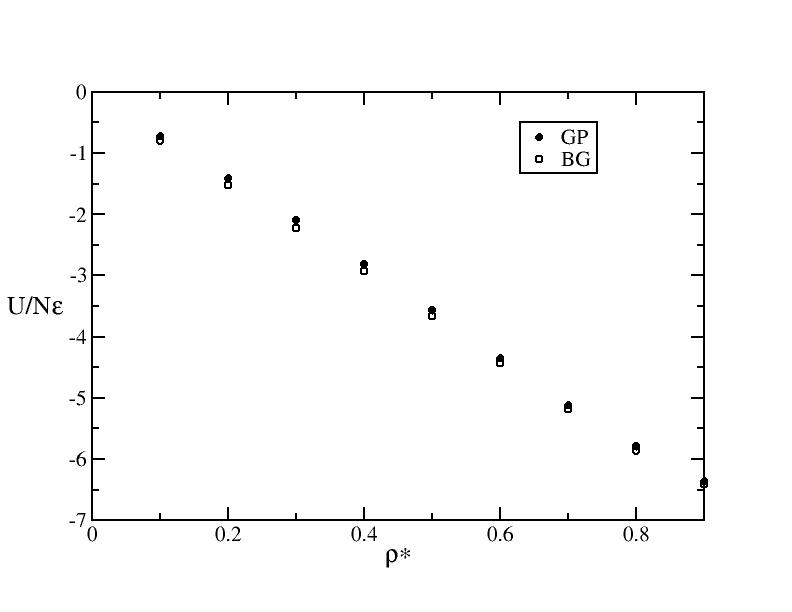}
\caption{NVT Monte Carlo simulation values for the Internal energy $U/N\epsilon$ for a system of
$N = 500$ spherical particles of diameter $\sigma$ interacting with
a square-well potential with energy depth $\epsilon$ and attractive range 
$\lambda = 1.5\sigma$. Results are presented for a supercritical temperature
$T^*  = 2.0$,
using the Boltzmann-Gibbs (BG) and generalized statistics (GP)
with triangle and circle symbols, respectively. 
Auxiliary interpolating lines are used as a guide for a better
visualization of the results.}
\label{fig:2}
\end{figure}

\begin{figure}[htb]
  \centering
    \includegraphics[scale=0.3]{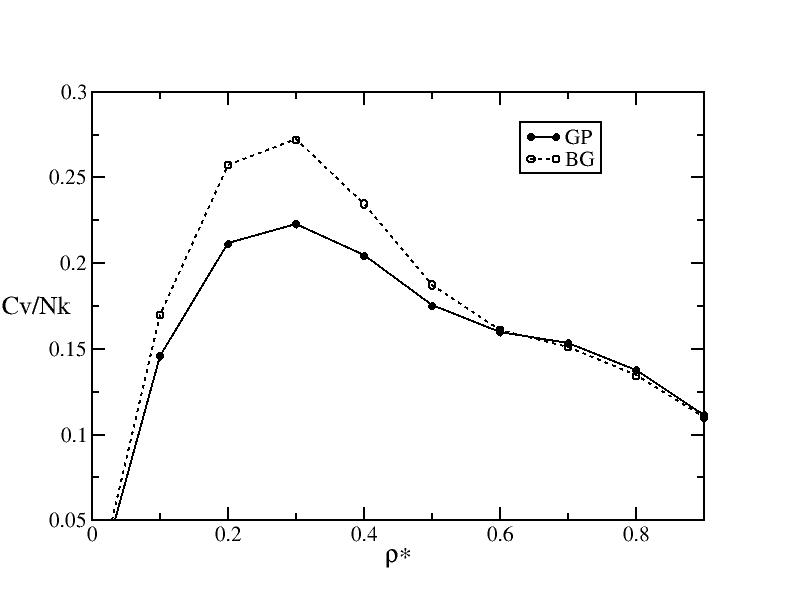}
\caption{NVT Monte Carlo simulation values for the constant-volume heat capacity  $C_V/Nk$ for a system of
$N = 500$ spherical particles of diameter $\sigma$ interacting with a square-well potential
with energy depth $\epsilon$ and attractive range 
$\lambda = 1.5\sigma$. Results are presented for a supercritical temperature
$T^* = 2.0$,
using the Boltzmann-Gibbs (BG) and generalized statistics (GP)
with black circle and open circle symbols, respectively. Auxiliary interpolating
lines are used as a guide for a better
visualization of the results.}
\label{fig:3}
\end{figure}

\begin{figure}[htb]
  \centering
    \includegraphics[scale=0.3]{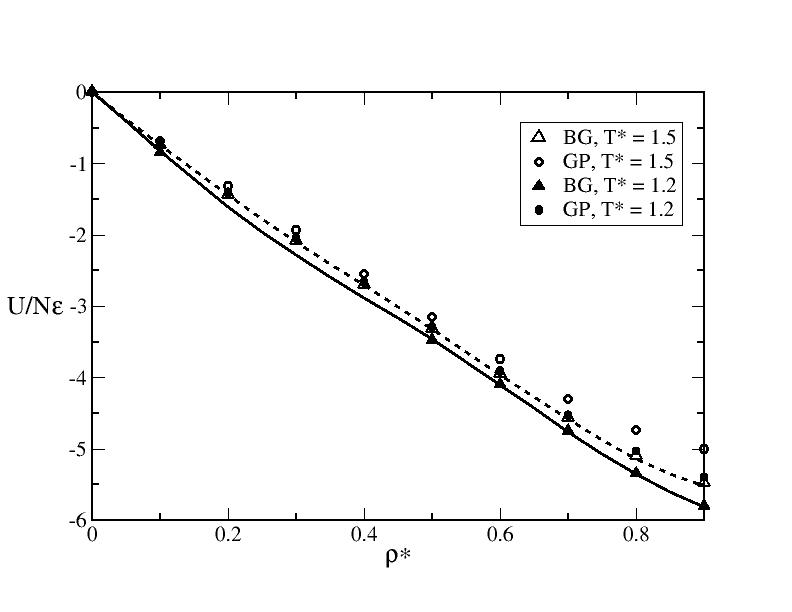}
\caption{NVT Monte Carlo simulation values for the internal energy
 $U/N\epsilon$ for a system of
$N = 864$ particles interacting with a Lennard-Jones potential
with energy depth $\epsilon$ and size parameter $\sigma$.  
Results are presented for a subcritical and a supercritical temperatures
$T^* = 1.2$ and $1.5$,
using the Boltzmann-Gibbs (BG) and generalized statistics (GP), as indicated in the figure.
For the BG cases we include the corresponding predictions using the Lennard-Jones
equation of state by Johnson {\it et al} in continuos and dashed  lines for $T^*=1.2$ and $T^*=1.5$ respectively.\cite{gubbins93}}
\label{fig:4}
\end{figure}

\begin{figure}[htb]
  \centering
    \includegraphics[scale=0.3]{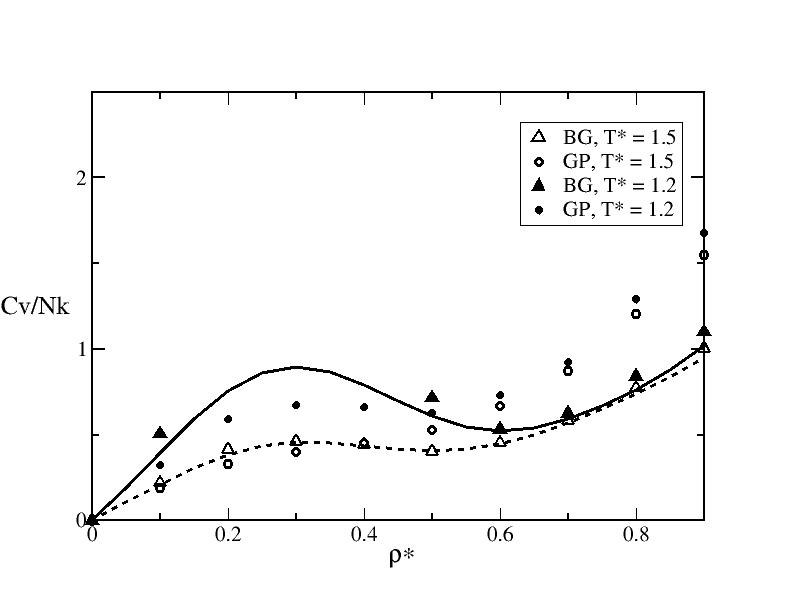}
\caption{NVT Monte Carlo simulation values for the constant-volume heat capacity
 $C_V/Nk$ for a system of
$N = 864$ particles interacting with a Lennard-Jones potential
with energy depth $\epsilon$ and size parameter $\sigma$.  
Results are presented for a subcritical and a supercritical temperatures
$T^* = 1.2$ and $1.5$,
using the Boltzmann-Gibbs (BG) and generalized statistics (GP), as indicated in the figure.
For the BG cases we include the corresponding predictions using the Lennard-Jones
equation of state by Johnson {\it et al.} in continuous and dashed lines for $T^*=1.2$ and $T^*=1.5$ respectively.\cite{gubbins93}}
\label{fig:5}
\end{figure}

In order to explore the consequences of redefining the distribution probability using a generalized
statistics, we performed canonical ensemble Monte Carlo computer simulations 
for a fluid composed of spherical particles interacting via 
two different models, square-well (SW) and Lennard-Jones (LJ) potentials.
In the first case we considered $N=500$ 
particles of diameter $\sigma$ 
interacting via a SW pair potential with an attractive range $\lambda = 1.5 \sigma$ and
energy-depth $\epsilon$; for the second model we used $N=864$ particles. 
Results were obtained for the internal energy $U/N\epsilon$ and the constant-volume heat capacity $C_V^*= C_V/Nk$,
using both
the standard Boltzmann-Gibbs statistics and the
generalized probability of this work \cite{6}, see Figures 2 and 3 for the SW system 
at temperature $T^*\equiv kT/\epsilon  = 2.0$, and Figures 4 and 5 for the LJ system at temperatures
$T^* = 1.2$ and $T^*= 1.5$.  In the first case we are studying a supercritical temperature, whereas
in the second case a comparison is made between a subcritical and supercritical case. 
In all the cases we observe that
 the generalized probability introduces an effective repulsive interaction. The effect on the thermodynamic properties is equivalent to increase
the repulsive force between molecules and there is a reduction on the values for the internal energy.
For the case of the heat capacity the effects of modifying the probability are more noticeable
near the critical density or at higher densities where clearly the GP values of $C_V*$  are greater for the Boltzmann-Gibbs statistics.
The procedure to define an effective potential $u_{eff}(r)$ is by assuming
that the generalized Boltzmann factor in equation (4) can be mapped onto
a classical Boltzmann factor,
\begin{equation}
e^{-\beta u_{eff}(r)} = B(E)
\end{equation}

An interesting consequence of this
effective-repulsive potential behavior is the possibility to map
the thermodynamic properties of a GP system onto a BG one by modifying the potential parameters, like 
the diameter $\sigma$ and range 
$\lambda$ of the SW fluid. This type of mapping has been explored in the past in order to define the
equivalence of thermodynamic properties between systems defined by different pair potentials \cite{gil96},
that now can be applied by mapping a GP and a BG potential-system.

If we think about the inverse problem, i.e, which is the
pair potential that reproduces the experimental values of $U/N\epsilon$ or $C_V/Nk$
of a real substance, it could be possible to assume that part of the information can be hindered
in the statistical probability function used to obtain the information,
i.e., to be considering either a BG statistics with a specific potential
model or a GP statistics with a modified pair potential.

\section{Conclusions}    

Based on a non-extensive statistical mechanics generalization of the  entropy that depends only on the probability,  we show that the first term correcting the usual entropy also arises from several $f(\beta)$  distributions.  We also construct the corresponding $H$-function and demonstrate that a 
 generalized $H$-theorem is fulfilled.  Furthermore, expressing this $H$ function as a function of the simplest Maxwellian state we find, up to a first approximation some  modified thermodynamic quantities for an ideal gas showing that a generic correction term appear, resembling the universal behavior founded for the distribution functions. Several simulation results are presented for internal energies and heat capacities for the square-well and Lennard-Jones potential. The simulation results support the theoretical results (for the ideal gas) showing that an effective repulsive interaction is obtained with the new formalism. Further research has to be done  to elucidate the complete scenario proposed 
 in this work.\\

\acknowledgments

We thank our supportive institutions. O. Obreg\'on was supported by CONACyT Projects No. 257919 and 258982, Promep and UG projects. J. Torres-Arenas was 
supported by CONACyT Project No. 152684 and Universidad de Guanajuato Project No. 740/2016.

\end{document}